\documentclass[letterpaper, 10 pt, conference]{ieeeconf}  % Comment this line out
\usepackage{comment}
\usepackage{graphics}
\usepackage{amsmath}
\usepackage{amssymb}
\usepackage{amsfonts}
\usepackage{cite}
\usepackage{adjustbox}
\usepackage{tikz}
\usetikzlibrary{positioning,arrows,shapes,chains,fit,scopes}
\usepackage{cite}
\usepackage{upgreek}
\usepackage{textcomp,float}
\usepackage{times} % assumes new font selection scheme installed
\usepackage{euscript}
\usepackage{color}
\usepackage{ccaption}
\usepackage{listings}
\usepackage{longtable}
\usepackage{multirow}
\usepackage{epstopdf}

\usepackage{makecell}
\usepackage{subfigure}
\usepackage{mathtools}
\usepackage{adjustbox}
\usepackage{array,esint}
\IEEEoverridecommandlockouts                              % This command is only
   \usepackage{booktabs}   
   \usepackage{caption}                                                    % needed if you want to
                                                         \usepackage{bbm}
\usetikzlibrary{calc}
\newcommand*\circled[1]{\tikz[baseline=(char.base)]{
    \node[shape=circle, draw, inner sep=1pt, 
        minimum height={\f@size*1.6},] (char) {\vphantom{WAH1g}#1};}}

\usepackage{calc}
\newtheorem{Theorem}{\bf \text{Theorem}}
\newtheorem{Definition}{\bf \text{Definition}}
\newtheorem{Proposition}{\bf \text{Proposition}}

\newtheorem{Lemma}[Theorem]{\bf \text{Lemma}}
\newtheorem{Remark}{\bf \text{Remark}}

\newtheorem{example}{\bf \text{Example}}

\let\Sigma\varSigma
\let\Gamma\varGamma

\makeatletter

\newcommand{\Rmnum}[1]{\expandafter \@slowromancap\romannumeral #1@}
\makeatother

\DeclareMathOperator*\argmin{arg\,min}

\DeclareFontFamily{U}{mathx}{}
\DeclareFontShape{U}{mathx}{m}{n}{<-> mathx10}{}
\DeclareSymbolFont{mathx}{U}{mathx}{m}{n}
\DeclareMathAccent{\widehat}{0}{mathx}{"70}
\DeclareMathAccent{\widecheck}{0}{mathx}{"71}

\title{\LARGE \bf
Kernel-Based Regularized Continuous-Time System Identification from Sampled Data
}

\author{Xiaozhu Fang, Biqiang Mu, and Tianshi Chen % <-this % stops a space
\thanks{This work was funded by NSFC under contract No. 62273287, Shenzhen Science and Technology Innovation Commission under contract No. JCYJ20220530143418040, the Thousand Youth Talents Plan funded by the central government of China, and the National Key R\&D Program of China under contract No. 2022YFA1004700.}% <-this % stops a space
\thanks{Xiaozhu Fang and Tianshi Chen are with School of Data Science and Shenzhen Research Institute of Big Data, The Chinese University of Hong Kong, Shenzhen, 518172 China (email: xiaozhufang@link.cuhk.edu.cn, tschen@cuhk.edu.cn) }%
\thanks{Biqing Mu is with the Key Laboratory of Systems and Control of
CAS, Institute of Systems Science, Academy of Mathematics and System
Science, Chinese Academy of Sciences, Beijing 100190, China (email: bqmu@amss.ac.cn)
}
}

\begin{document}

\maketitle
\thispagestyle{empty}
\pagestyle{empty}

%%%%%%%%%%%%%%%%%%%%%%%%%%%%%%%%%%%%%%%%%%%%%%%%%%%%%%%%%%%%%%%%%%%%%%%%%%%%%%%%
\begin{abstract}
The identification of continuous-time (CT) systems from discrete-time (DT) input and output signals, i.e., the sampled data, has received considerable attention for half a century. The state-of-the-art methods are parametric methods and thus subject to the typical issues of parametric methods. In the last decade, a major advance in system identification is the so-called kernel-based regularization method (KRM), which is free of the issues of parametric methods.  It is interesting to test the potential of KRM on CT system identification. However, very few results have been reported, mainly because the estimators of KRM have no closed forms for general CT input signals, except for some very special cases. In this paper, we show for KRM that the estimators have closed forms when the DT input signal has the typical intersample behavior, i.e.,  zero-order hold or band-limited, and this paves the way for the application of KRM for CT system identification. Numerical simulations show that the proposed method is more robust than the state-of-the-art methods and more accurate when the sample size is small.   
\end{abstract}
%%%%%%%%%%%%%%%%%%%%%%%%%%%%%%%%%%%%%%%%%%%%%%%%%%%%%%%%%%%%%%%%%%%%%%%%%%%%%%%%
\section{INTRODUCTION}
Continuous-time (CT) systems have received increasing interest due to their association with the natural world and physical processes through derivatives, integrals, and differential equations. 
% The sensors of CT systems are frequently accessed from digital electronics, suggesting that the input and output are the sampled signals, i.e. they are discretized by sampling interval and windowed by sample size from the continuous signals. 
The identification of CT systems from the discrete-time (DT) input and output signals (which are also called sampled data) has been a long-standing topic for half a century, see survey papers \cite{Young:81, UR90, Garnier15}. The existing methods can be divided into two classes: indirect ones and direct ones, see e.g., \cite{UR90, Garnier15}. Specifically, the direct methods identify a CT parametric model directly from the sampled data, while the indirect methods first identify a DT model from the sampled data, and then convert it to a CT parametric model. The choice between direct methods and indirect methods has been controversial for decades, and recently, the mainstream has been the direct methods. Particularly, the state-of-the-art direct methods are the Simplified Refined Instrumental Variable method for Continuous-time system  (SRIVC) \cite{YJ80,GLY08}, and the Maximum Likelihood/Prediction Error Method (ML/PEM) \cite{Ljung:99, LJUNG09}. However, due to the use of parametric models, both SRIVC and ML/PEM are subject to the typical issues of parametric methods. First, the model order should be decided in advance, and the problem of finding a suitable model order is challenging, especially when the data is short and/or has a low signal-to-noise ratio. Second, when estimating a model with a large order,  the nonconvex optimization problem involved in the parameter estimation often has severe local minima issues, leading to large variations between the different noise realizations, see \cite[Sec. 6.5]{LJUNG09}.

% The DT input signal is insufficient to build the infinite impulse response model, and extra knowledge of the input behavior is necessary to know.   On the one hand, it is necessary to know the intersample behavior of the input signal, e.g. zero-order-hold (ZOH) and band-limited (BL), and the misuse of the intersample behavior would affect the estimator's consistency, see e.g.\cite{SPV94, Garnier15, GRPW21}.  On the other hand, for infinite impulse response, the past input is needed but unknown; as a result, it is commonly assumed that the input signal has past behavior such as periodically appended (PA) or zero appended (ZA) with the additionally modeling the transient error (also known as the initial condition error), see e.g. \cite{Ljung04,PS12}.  Given the intersample and past behavior, the CT input signal can be reconstructed from the sampled input signal. 

In the last decade, a major advance in system identification is the so-called kernel-based regularization method (KRM). It is free of the aforementioned issues of parametric methods, and moreover, extensive numerical and experimental simulation results have shown that, for DT system identification, KRM can provide more accurate and robust DT model estimates than ML/PEM when the data is short and/or has a low signal-to-noise ratio, and KRM has become an emerging paradigm in system identification \cite{LCM20}. The success of KRM in DT system identification naturally encourages researchers to further explore whether such success can be copied in CT system identification. There have been some attempts recently, see e.g. \cite{PDCDL14,SMFP22}.  However, the integral involved in the computation of the predicted output has no closed form, leading to the absence of the closed form of the estimators. An accurate approximation of the integral is feasible but expensive, and thus KRM has not been essentially tested for CT system identification, except for some very special cases, e.g. the input is an impulse or step signal \cite{PD10,SMFP22}.

In this paper, we apply KRM to CT system identification from the sampled data with extra knowledge of the input behaviour.  Specifically, we convert the DT input signal into a CT input signal by exploiting the given intersample behavior, e.g. zero-order-hold (ZOH) and band-limited (BL). 
Moreover, KRM also needs to access to the past input, which is unknown in the sampled data; as a result, the past behavior of the input signal needs to be assumed, e.g. periodically appended (PA) or zero appended (ZA), and the inconsistency of this assumption is also modeled as the transient error, see e.g. \cite{Ljung04,PS12}. Consequently, KRM can be adopted for the sampled data, and notably, we have found that the closed forms of the estimators exist for certain combinations of the intersample and past behaviors: ZOH and PA, ZOH and ZA, BL and PA, which enables the efficient and accurate implementation of KRM. Lastly, we run numerical simulations to compare the proposed methods with SRIVC and ML/PEM, where the proposed method not only has better robustness but also, when the sample size is small, has higher accuracy.

%The GPR to the SDSI is a special case of the GPR for the CTSI\cite{PDCDL14} via using the intesample and past behavior. The CFE without the closed forms needs to be approximated by means of numerical integration techniques \cite{DINUZZO15}, which is inaccurate and computationally inefficient. 
%In this paper, however, we show that CFE exists for certain combinations of the intesample and past behavior, i.e., the input signal is ZOH and PA, ZOH and ZA, or BL and PA, enabling the efficient implementation of the GPR.  Moreover, the result 
%is extended to the case when the input signal is ZOH or BL but has unknown past behavior. Lastly, numerical simulations are run to compare the GPR with the SRIVC and ML/PEM for the ZOH input.   
The remaining part of this paper is organized as follows. In Section \ref{se:problem}, CT system identification and KRM  are briefly reviewed. In Section \ref{se:sd_id}, KRM is applied to CT system identification from the sampled data, and the results are also extended to the unknown past behavior. Lastly, numerical simulations are shown in Section \ref{se:sim}. The proofs of the propositions and lemmas are skipped unless otherwise stated. 
\section{Background and Problem Statement}\label{se:problem}
\subsection{CT System Identification from the Sampled Data}\label{se:sdsi}
Consider a bounded-input-bounded-output (BIBO) stable, strictly causal, continuous-time (CT), single-input single-output (SISO), and linear time-invariant (LTI) system,  described by
{\small\begin{align}\label{eq:sys_ir}
&y(t) = y_0(t)+ v(t), \ y_0(t) =  \int_0^\infty u(t-\tau)g(\tau)d\tau, 
\end{align}}\normalsize
where $t, \tau\in \mathbb{R}$ are time indexes, $u(t), y(t)\in \mathbb{R}$ are the input and  output of the system, respectively, $v(t)\in \mathbb{R}$ is the disturbance,  $g(\tau)\in \mathbb{R}$ is the impulse response of the system, and $y_0(t)\in \mathbb{R}$ is the noiseless output, i.e., the
convolution between the impulse response $g(\cdot)$ and the input $u(\cdot)$ evaluated at time $t$.

Let $u(t)$ and $y(t)$ sampled at $t=0,T_s, \cdots, (N-1)T_s$, where $T_s$ is the sampling interval. 
Since $y(t)$ is sampled, the disturbance $v(t)$ can be characterized by a discrete-time (DT) model \cite[Sec. 2.2]{GY14}, and, moreover,  $v(t)$ is assumed to be white Gaussian distributed with mean zero and variance $\sigma^2$,  independent of $u(t)$.

Our goal is to identify the CT impulse response $g(\tau)$ from the sampled data, i.e., $N$ sampled measurement of the input and output $\{u(kT_s), y(kT_s)\}_{k=0}^{N-1}$. 
\subsection{Kernel-Based Regularization Method}\label{se:bayesian_ctsi}
Recently, the kernel-based regularization method (KRM) has been intensively discussed in the context of system identification. KRM has been applied to CT system identification from the CT input signal $\{u(t)\}_{t\leq NTs}$  and DT output signal  $\{ y(kT_s)\}_{k=0}^{N-1}$, see e.g.  \cite{PDCDL14,SMFP22}. 
 Specifically, the CT impulse response $g$ is estimated by the regularized Least Squares as
{\small \begin{align}
& \hat{g}\! =\! \underset{g\in \mathcal{H}_{g}}{\argmin}  \sum\limits_{k=0}^{N-1}\bigg|y(kT_s)\!-\int_0^\infty u(kT_s-\tau)g(\tau)d\tau\bigg|^2\!+\!\gamma \|g\|^2_{\mathcal{H}_{g}}\label{eq:rls}
\end{align}  }\normalsize
where $\gamma\geq 0$ is the regularization parameter, and $\mathcal{H}_{g}$ is a reproducing kernel Hilbert spaces   induced by the kernel $\kappa_g: \mathbb{R}\times  \mathbb{R}\rightarrow  \mathbb{R}$. It is well-known that $\hat{g}$ in \eqref{eq:rls} can be obtained from the representer theorem \cite{PDCDL14}, leading to the following matrix-vector form: 
{\small\begin{align}
&\hat{g}(\tau)= \pmb  \Sigma_{gy_0}(\tau) (\pmb \Sigma_{y_0}+ \gamma I_N)^{-1}\pmb y, \label{eq:map_ir} \\
&\left\{\begin{array}{lll}
\pmb \Sigma_{y_0}\!=\!
\int_{0}^\infty \int_{0}^\infty  u(\pmb t\!-\!\tau) u(\pmb t^T\!-\!\tau')\kappa_{g}(\tau,\tau')\,d\tau \,d \tau' ,  \\
\pmb \Sigma_{gy_0}(\tau)\!=\!
 \int_0^\infty  u(\pmb t^T-\tau')\kappa_{g}(\tau,\tau') d\tau', 
 \end{array}\right.\label{eq:map_y}
\end{align}}\normalsize
where $\pmb y=[y(0),\cdots, y((N-1)T_s)]^T$, $I_N$ is a $N\times N$ identity matrix, $\pmb t=[0,\cdots, (N-1)T_s]^T$ with  $\pmb t^T$ denoting its transpose, and the notation of vectorized scalar functions is employed above\footnote{Given a scalar function $\xi(\cdot)$,  the vectorized scalar function $\xi(\pmb t)$  outputs the vector $[\xi(0),\cdots, \xi((N-1)T_s)]^T$.  Similarly, this notation can be used for the scalar function with two vector inputs, e.g.,$
\xi(\pmb t,\pmb t^T)$,  which outputs a matrix.}.

It should be noted that the solution \eqref{eq:map_ir} can also be interpreted as a maximum a posteriori (MAP) estimate in the following Bayesian framework. 
\begin{Lemma}[\cite{PDCDL14}]\label{le:rels_time}
Consider \eqref{eq:sys_ir} with:
\begin{enumerate}
\item $g(\tau)$ is a zero-mean Gaussian process with covariance function $\kappa_g(\tau,\tau')$,
\item $v(t)$ is a white Gaussian noise with variance $\sigma^2= \gamma$, independent with  $g(\tau)$.
\end{enumerate}
Given observation $\pmb y=[y(0),\cdots, y((N-1)T_s)]^T$, 
the MAP estimate of $g(\tau)$ coincides with \eqref{eq:map_ir}.
\end{Lemma}

It should be noted that the estimator \eqref{eq:map_ir} has the matrix-vector form, and then \eqref{eq:map_ir} has the closed form if  matrices $\pmb{\Sigma}_{gy_0}(\tau)$  and $\pmb{\Sigma}_{y_0}$ have closed forms. Nonetheless, for general $u(t)$, matrices $\pmb{\Sigma}_{gy_0}(\tau)$  and $\pmb{\Sigma}_{y_0}$ have no closed form due to the infinite-dimensional integral.  Though numerical integral techniques can be employed,   the computation  would be either inaccurate or expensive.  Thus, the absence of the closed forms of $\pmb{\Sigma}_{gy_0}(\tau)$  and $\pmb{\Sigma}_{y_0}$ has been the main barrier of the applications of KRM. 
\begin{Remark}
The closed form of $\pmb{\Sigma}_{y_0}$ also influences the hyperparameter estimation, e.g. the
marginal likelihood \cite{PDCDL14}, in the same manner.
\end{Remark}

\subsection{Problem Statement}
In what follows, we consider KRM for CT system identification from the sampled data. The idea is straightforward by exploiting \eqref{eq:rls}, but the underlying input is DT signal  $\{u(kT_s)\}_{k=0}^{N-1}$ rather than CT signal $\{u(t)\}_{t\leq  NT_s}$. 
Then the main problem can be addressed as follows:  
\begin{itemize}
\item How to construct $\{u(t)\}_{t\leq  NT_s}$ from $\{u(kT_s)\}_{k=0}^{N-1}$ to obtain the regularized estimator \eqref{eq:map_ir}, and moreover, the closed form of $\pmb{\Sigma}_{gy_0}(\tau)$  and $\pmb{\Sigma}_{y_0}$? 
\end{itemize}
The answer will be given in Section \ref{se:sd_id}. 
 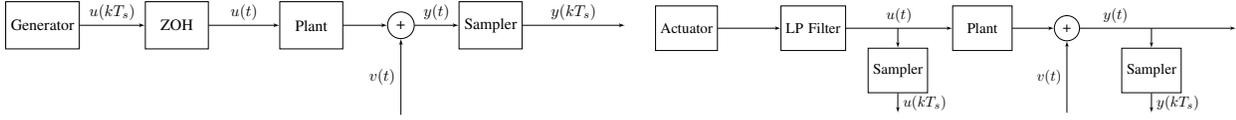
\begin{figure*}
   \resizebox{0.48\textwidth}{!}{%
\tikzstyle{block} = [draw, rectangle, 
    minimum height=3em, minimum width=4em]
\tikzstyle{sum} = [draw, circle, node distance=1cm]
\tikzstyle{dio} = [draw, diamond, node distance=1cm]
\tikzstyle{pinstyle} = [pin edge={to-,thin,black}]
\begin{tikzpicture}[auto, node distance=2cm,>=latex']
    \node [coordinate, name=input] {};
    \node [block, right of=input, node distance=3cm] (t1) {Generator};      
    \node [block, right of=t1, node distance=3cm] (t2) {ZOH};
        \node [block, right of=t2, node distance=3cm] (t3) {Plant 
};
    \node [coordinate, right of=t3, node distance=3cm] (t4) {};
        \node [sum, right of=t3, node distance=2cm] (t4) {+};
        \node [coordinate, below of=t4, node distance=2cm] (s4) {};
             \node [block, right of=t4, node distance=2cm] (t5) {Sampler};
                          \node [coordinate, right of=t5, node distance=3cm] (t6) {};
  	     	\draw [->] (t1) -- node {$u(kT_s)$} (t2);
  	     	\draw [->] (t2) -- node {$u(t)$} (t3);
       	 	\draw [->]  (t3) -- node {} (t4);
       	 	 	 	\draw [->]  (s4) -- node {$v(t)$} (t4);
       	 	       	 	\draw [->]  (t4) -- node {$y(t)$} (t5);
       	 	       	 	       	 	\draw [->]  (t5) -- node {$y(kT_s)$} (t6);
\end{tikzpicture}}
 \resizebox{0.45\textwidth}{!}{%
\tikzstyle{block} = [draw, rectangle, 
    minimum height=3em, minimum width=4em]
\tikzstyle{sum} = [draw, circle, node distance=1cm]
\tikzstyle{dio} = [draw, diamond, node distance=1cm]
\tikzstyle{pinstyle} = [pin edge={to-,thin,black}]
\begin{tikzpicture}[auto, node distance=2cm,>=latex']
    \node [coordinate, name=input] {};
    \node [block, right of=input, node distance=3cm] (t1) {Actuator};      
    \node [block, right of=t1, node distance=3cm] (t2) {LP Filter};
      \node [coordinate, right of=t2, node distance=2cm] (t25) {};  
        \node [block, right of=t25, node distance=2cm] (t3) {Plant 
  };
    \node [coordinate, right of=t3, node distance=3cm] (t4) {};
        \node [sum, right of=t3, node distance=2cm] (t4) {+};
                          \node [block, below of=t25, node distance=1cm] (r25) {Sampler};
                                         \node [coordinate, below of=r25, node distance=1cm] (p25) {};
        \node [coordinate, below of=t4, node distance=2cm] (s4) {};
             \node [coordinate, right of=t4, node distance=2cm] (t5) {};
               \node [block, below of=t5, node distance=1cm] (r5)  {Sampler};
                     \node [coordinate, below of=r5, node distance=1cm] (p5)  {};
  	     	\draw [->] (t1) -- node {} (t2);
  	     	\draw [->] (t2) -- node {$u(t)$} (t3);
  	     	 	     	 	 	     	\draw [->] (t25) -- node {} (r25);
  	     	 	     	 	 	     	   	\draw [->] (r25) -- node {$u(kT_s)$} (p25);
       	 	\draw [->]  (t3) -- node {} (t4);
       	 	 	 	\draw [->]  (s4) -- node {$v(t)$} (t4);
       	 	       	 	\draw [-]  (t4) -- node {$y(t)$} (t5);
       	 	       	 	       	 	\draw [->]  (t5) -- node {} (t6);
  	     	 	     	 	 	     	\draw [->] (t5) -- node {} (r5);
  	     	 	     	 	 	     	   	\draw [->] (r5) -- node {$y(kT_s)$} (p5);
\end{tikzpicture}}

\caption{Basic setup for ZOH (left) and BL (right) inputs in applications, where $u(kT_s)$ and $y(kT_s)$ are measured, and LP Filter denotes the low-pass filter for BL conditions. Refers the more general setup to \cite[Chap. 13.2]{PS12}.  }\label{fig:blk_sys}
\end{figure*}

\section{Main Result}\label{se:sd_id}
In what follows, we first revisit how to construct $\{u(t)\}_{t\leq  NT_s}$ from $\{u(kT_s)\}_{k=0}^{N-1}$  by using the input intersample and input past behaviors in Section \ref{se:bay_input}.  Then, given such behaviors,   we discuss the closed forms of $\pmb{\Sigma}_{gy_0}(\tau)$  and $\pmb{\Sigma}_{y_0}$  in Section \ref{se:cf_known}. For practical use, the results are extended to the unknown past behavior in Section \ref{se:cf_unknown}. 
\subsection{From Sampled Input Signal to CT Input  Signal}\label{se:bay_input}
The construction of $\{u(kT_s)\}_{k=0}^{N-1}$ is considered in two ways. 
On the one hand, the DT input signal needs to be converted into a CT input signal by using intersample behavior \cite{SPV94}. The typical intersample behaviors are  
\begin{itemize}
\item  zero-order hold (ZOH), i.e., 
{\small\begin{align}
&u(t)=  \sum\limits_{k=0}^\infty u(kT_s) zoh(t-kT_s) \text{ with}\nonumber\\
&zoh(t)= \mathrm{1}_{(0,\infty)}(t)-\mathrm{1}_{(T_s,\infty)}(t)\label{eq:zoh_assupt}, 
\end{align}}\normalsize
where $\mathrm{1}_{\ast}(t)$ is the indicator function, i.e., equal 1 when condition $\ast$ is satisfied.  
\item  band-limited (BL) under the Nyquist frequency, i.e.,  containing no frequencies on and above $\pi/T_s$ rad/s.  Further, if $u(t)$ is periodic with period $NT_s$ and BL under  $\pi/T_s$, then it holds that
{\small\begin{align*}
u(t)= \frac{1}{N}\sum\limits_{|n|<\frac{N}{2}}\sum\limits_{k=0}^{N-1}u(kT_s)e^{-j 2\pi nk/N } e^{j2\pi nt/NT_s}, 
\end{align*}}\normalsize
\end{itemize} 
The intersample behavior is based on the measurement setup, see Fig.\ref{fig:blk_sys},  and thus it is assumed known in this paper.

On the other hand, the past input, $u(t)$ with $t<0$, needs to be specified in \eqref{eq:map_y}, and it can be inferred by assuming the past behavior \cite[Sec. VIII]{Ljung04}. The typical past behaviors are
\begin{itemize}
\item periodically appended (PA) with period $NT_s$, i.e., 
{\small\begin{align}
u(t-kNT_s)=u(t), \quad  0\leq t<NT_s, k\in\mathbb{N}, \label{eq:per_assupt}
\end{align}}\normalsize
where PA in this paper  shall always carry the information of periods $NT_s$;
\item zero appended (ZA), i.e., 
{\small\begin{align}
u(t)= 0, \text{ when } t<0, \label{eq:zero_assupt}
\end{align}}\normalsize
which is also known as the per-windowing approach  \cite[p.320]{Ljung:99}. 
\end{itemize}
The past behavior is often unknown, and the assumed PA and ZA behaviors are very likely to be inconsistent,  leading to the estimation error. This error can be modeled in addtition as the transient in Section \ref{se:cf_unknown}.   For the moment, nonetheless, we assume the known intersample and past behaviors.

\subsection{Closed Forms for Known Intersample and Past Behaviors}\label{se:cf_known}

Given the intersample and past behaviors,  $\{u(t)\}_{t\leq  NT_s}$ can be constructed from $\{u(kT_s)\}_{k=0}^{N-1}$, and thus KRM can be applied. 
Then we consider the closed forms of $\pmb{\Sigma}_{gy_0}(\tau)$  and $\pmb{\Sigma}_{y_0}$ for different intersample and past behaviors. 
Since BL is contradictory to ZA, there are three different combinations to be considered: ZOH and PA, ZOH and ZA, BL and PA.
 
\subsubsection{ZOH and PA Input}\label{se:dint}
We first consider the ZOH input and PA input, respectively, and then their combination. On the one hand, if $u(t)$ is ZOH, then the CT impulse response $g(\tau)$ can be replaced by a DT model, see e.g. \cite{SPV94}. This leads to the following lemma. 
\begin{Lemma}[ZOH Input]\label{pr:zoh_input}
If $u(t)$ is ZOH, then 
 \eqref{eq:map_y} can be rewritten as 
{\footnotesize\begin{align}\label{eq:zoh_y_time} 
&\left\{
\begin{array}{lll}
\pmb {\Sigma}_{y_0}\!=\!   \sum\limits_{s =1}^\infty   \sum\limits_{s' =1}^\infty   u(\pmb t-sT_s)u(\pmb t^T-s'T_s) \kappa_{g_d}(sT_s, s'T_s),\\
\pmb {\Sigma}_{gy_0}(\tau)\!=\!    \sum\limits_{s' =1}^\infty u(\pmb t^T-s'T_s) \kappa_{gg_d}(\tau, s'T_s), 
\end{array}\right.\\
& \kappa_{g_d}(sT_s, s'T_s)\!= \!\int_{(s-1)T_s}^{sT_s}\int_{(s'-1)T_s}^{s'T_s} \kappa_g (\tau, \tau')\, d\tau\, d\tau', \nonumber\\
&\kappa_{gg_d}(\tau, s'T_s)\!= \! \int_{(s'-1)T_s}^{s'T_s} \kappa_g (\tau, \tau')\, d\tau',\quad  s,s'\in \mathbb{N}_+. \nonumber
\end{align}}\normalsize
\end{Lemma}
 On the other hand, if $u(t)$ is PA, then the CT impulse response $g(\tau)$ can be replaced by a truncated CT model. This leads to the following lemma. 
\begin{Lemma}[PA Input]
If $u(t)$ is PA with period $NT_s$, then 
 \eqref{eq:map_y} can be rewritten as 
{\footnotesize\begin{align}\label{eq:zoh_y_N}
&\left\{\begin{array}{lll}
\pmb \Sigma_{y_0}\!=\! 
\int_{0}^{NT_s} \int_{0}^{NT_s} u(\pmb t-\tau) u(\pmb t^T-\tau') \kappa_{g_p}(\tau,\tau')\,d\tau \,d \tau',\\
\pmb \Sigma_{gy_0}(\tau)\!=\!
 \int_0^{NT_s}  u(\pmb t^T-\tau')\kappa_{gg_p}(\tau,\tau') d\tau', 
 \end{array}\right.\\
& \kappa_{g_p}(\tau, \tau')\!=\! \sum\limits_{n=0}^\infty\sum\limits_{n'=0}^\infty \kappa_{g}(\tau+ nNT_s, \tau'+n'NT_s), \nonumber\\
&\kappa_{gg_p}(\tau, \tau')\!=\! \sum\limits_{n'=0}^\infty \kappa_{g}(\tau, \tau'+n'NT_s), \quad 0 \leq \tau, \tau<  NT_s. \nonumber
\end{align}}\normalsize
\end{Lemma}
However, from \eqref{eq:zoh_y_time} and \eqref{eq:zoh_y_N},  relying alone on ZOH or PA does not suffice the closed forms of  $\pmb{\Sigma}_{gy_0}(\tau)$  and $\pmb{\Sigma}_{y_0}$. Then we consider the combination of them. 
If $u(t)$ is PA and ZOH, then the CT impulse response $g(\tau)$ can be replaced by a truncated DT model, leading to the following lemma.
\begin{Lemma}[ZOH and PA Input]\label{le:per_zoh_input}
If $u(t)$ is ZOH and PA with period $NT_s$, then 
 \eqref{eq:map_y} can be rewritten as 
{\footnotesize\begin{align}\label{eq:zoh_y_d_N}
& \left\{
\begin{array}{lll}
\pmb {\Sigma}_{y_0}\!=\!    \sum\limits_{s =1}^N   \sum\limits_{s' =1}^N   u(\pmb t-sT_s)u(\pmb t^T-s'T_s) \kappa_{g_{dp}}(sT_s, s'T_s), \\
\pmb {\Sigma}_{gy}(\tau)\!=\!    \sum\limits_{s' =1}^N u(\pmb t^T-s'T_s) \kappa_{gg_{dp}}(\tau, s'T_s), 
\end{array}\right.\\
& \kappa_{g_{dp}}(s, s')\!=\!\sum\limits_{n=0}^\infty\sum\limits_{n'=0}^\infty \kappa_{g_d}(sT_s+ nNT_s, s'T_s+n'NT_s), \nonumber\\
&\kappa_{gg_{dp}}(\tau, s')= \sum\limits_{n'=0}^\infty \kappa_{gg_d}(\tau, s'T_s+n'NT_s), \nonumber
\end{align}}\normalsize
where  $\kappa_{g_{dp}}$ and $\kappa_{gg_{dp}}$ can also induced from $\kappa_{g_p}$ and $\kappa_{gg_p}$. 
\end{Lemma}
From Lemma \ref{le:per_zoh_input}, it is shown that the closed forms of  $\pmb{\Sigma}_{gy_0}(\tau)$  and $\pmb{\Sigma}_{y_0}$ exist as long as the $\kappa_{g_{dp}}$ and $\kappa_{gg_{dp}}$  have the closed forms. Nonetheless, their closed forms depend on the selection of the original kernel $\kappa_g$. Particularly, we consider a widely-used kernel for $\kappa_g$  as an example. 
\begin{example}\label{ex:dc_kernel}
If  $\kappa_g(\tau, \tau')$ takes the form of the diagonal correlated (DC) kernel \cite{COL12}, described by 
{\footnotesize\begin{align}
   & \kappa_g^{\text{DC}}(\tau,\tau')=\lambda e^{-\alpha  (\tau+\tau')} e^ {-\beta  | \tau-\tau'| },\label{eq:dc_kernel_ct}
\end{align} }\normalsize
where $\alpha, \beta, \lambda$ are hyperparameters. Then we have 
{\footnotesize \begin{align*}
  \bullet\  &\kappa_{g_d}^{\text{DC}}(sT_s,s'T_s) = \left\{
  \begin{array}{lll}
      \lambda_1\lambda e^{-\beta |s-s'|T_s-\alpha (s +s' )T_s}, &\text{ if } s\neq s'\\
      \lambda_2\lambda e^{-\alpha(s+ s')T_s } , &\text{ if } s=s'
  \end{array}
\right. , \nonumber\\    
    & \lambda_1 = \frac{\left(1-e^{(\alpha-\beta)T_s }\right) \left(1-e^{(\alpha +\beta)T_s }\right) }{ \alpha^2 -\beta^2 },\nonumber\\
    &\lambda_2 =\frac{ -2 \alpha  e^{(\alpha -\beta)T_s }-e^{2 \alpha T_s } (\beta -\alpha )+\alpha +\beta }{ \alpha^3 -\alpha \beta^2 } \nonumber ,\\
     \bullet\  &\kappa_{g_p}^{\text{DC}}(\tau,\tau') = \lambda_3\lambda e^{ -\alpha  (\tau+ \tau')-\beta  |\tau-\tau'|}  + \lambda_3\lambda_4\lambda e^{-\alpha  (\tau+\tau')+\beta  (\tau-\tau')}\\
     & \qquad  + \lambda_3\lambda_4\lambda e^{-\alpha  (\tau+\tau')+\beta  (\tau'-\tau)}\\
&\lambda_3= \frac{ 1}{1- e^{-2 \alpha  NT_s}},   \quad \lambda_4= \frac{ e^{-\beta NT_s-\alpha NT_s}}{1-e^{-\alpha NT_s -\beta NT_s}},    \\
  \bullet\  &\kappa_{g_{dp}}^{\text{DC}}(s T_s,s' T_s) =  \lambda_3\kappa_{g_d}^{\text{DC}}(sT_s,s'T_s)+ \lambda_1\lambda_3\lambda_4\lambda   e^{-\alpha  (s+s')T_s+\beta  (s-s')T_s}  \\
  & + \quad \lambda_1\lambda_3\lambda_4\lambda   e^{-\alpha  (s+s')T_s+\beta  (s'-s)T_s} ,
\end{align*}}\normalsize
where the closed forms of $\kappa_{gg_{d}}$, $\kappa_{gg_{p}}$, and $\kappa_{gg_{dp}}$ are left in Appendix \ref{ap:dc_example}. 
\end{example}
From Example \ref{ex:dc_kernel}, we have the following proposition. 
\begin{Proposition}\label{pr:zoh_pc_input}
If the input signal is ZOH and PA,  and if $\kappa_g(\tau, \tau')$  is the DC kernel \eqref{eq:dc_kernel_ct}, then the closed forms of  $\pmb{\Sigma}_{gy_0}(\tau)$  and $\pmb{\Sigma}_{y_0}$ exist as \eqref{eq:zoh_y_d_N}. 
\end{Proposition}

\subsubsection{ZOH and ZA Input}
 If $u(t)$ is ZA, then the CT impulse response $g(\tau)$ can be truncated to $NT_s$, i.e.,  
{\small\begin{align}
&y_0(t)= \int_{0}^{NT_s}  u(t-\tau)g(\tau)\,d \tau, \label{pr:y_zc}
\end{align}}\normalsize
and moreover,  If $u(t)$ is  ZOH and ZA, then the DT model can be truncated to $N$, leading to the following lemma. 
\begin{Lemma}[ZOH and ZA Input]
If $u(t)$ is ZOH and ZA, then 
 \eqref{eq:map_y} can be rewritten as 
{\footnotesize\begin{align}\label{eq:zoh_zc_y_time} 
&\left\{
\begin{array}{lll}
\pmb {\Sigma}_{y_0}\!=\!   \sum\limits_{s =1}^N  \sum\limits_{s' =1}^N  u(\pmb t\!-\!sT_s)u(\pmb t^T\!-\!s'T_s) \kappa_{g_d}(sT_s, s'T_s),\\
\pmb {\Sigma}_{gy_0}(\tau)\!=\!    \sum\limits_{s' =1}^N u(\pmb t^T\!-\!s'T_s) \kappa_{gg_d}(\tau, s'T_s), 
\end{array}\right.
\end{align}}\normalsize
where  $\kappa_{g_d}$ and  $\kappa_{gg_d}$ refer to \eqref{eq:zoh_y_time} .
\end{Lemma}
\begin{Proposition}\label{pr:zoh_zc_input}
If the input signal is ZOH and ZA, and if $\kappa_g(\tau, \tau')$  is the DC kernel \eqref{eq:dc_kernel_ct}, then the closed forms of  $\pmb{\Sigma}_{gy_0}(\tau)$  and $\pmb{\Sigma}_{y_0}$ exist as \eqref{eq:zoh_zc_y_time}. 
\end{Proposition}

\subsubsection{BL and PA Input}
If $u(t)$ is BL and PA, then the corresponding result is complicated as summarized in the following lemma. 
\begin{Lemma}[PA Input and BL Input]\label{le:zoh_bl_input}
If $u(t)$ is BL and PA, then 
 \eqref{eq:map_y} can be rewritten as 
{\footnotesize
\begin{align}\label{eq:zoh_y_bl} 
& \left\{
\begin{array}{lll}
\pmb {\Sigma}_{y_0}\!=\!   \frac{1}{N^2}  \sum\limits_{|n|< \frac{N}{2}} \sum\limits_{|n'|< \frac{N}{2}}  e^{jn \omega_0  \pmb{t}}e^{-jn' \omega_0  \pmb{t}^T} \quad U(n\omega_0) \overline{U}(n'\omega_0)\\
\qquad   \int_{0}^{NT_s}\int_{0}^{NT_s} \kappa_{g_p}(\tau, \tau')  e^{-j n\omega_0 \tau}  e^{j n'\omega_0 \tau'}  \,d \tau\,d \tau' , \\
\pmb {\Sigma}_{gy}(\tau)\!=\!  \frac{1}{N}  \sum\limits_{|n'|< \frac{N}{2}}  e^{j n' \omega_0  \pmb{t}^T}\overline{U}(n'\omega_0 )
 \int_{0}^{NT_s} \kappa_{gg_p} (\tau, \tau’)  e^{j n'\omega_0 \tau’}  \,d \tau’, 
\end{array}\right.
\end{align}}\normalsize
where $\omega_0= 2\pi/NT_s$,  $U(n\omega_0)= \sum_{k=0}^{N-1} u(kT_s) e^{-jn\omega_0kT_s}$, and  $\overline{U}(n\omega_0)$ is  the complex conjugate of $U(n\omega_0)$. 
\end{Lemma}
 Likewise to Example \ref{ex:dc_kernel}, if we choose the DC kernel \eqref{eq:dc_kernel_ct}, then $\pmb {\Sigma}_{y_0}$ and  $\pmb {\Sigma}_{gy}(\tau)$ also have the closed form, but we skip it for limited space. 
Lastly, we have the following proposition. 
\begin{Proposition}\label{pr:bl_pc_input}
If the input signal is BL and PA, and if  $\kappa_g(\tau, \tau')$  is the DC kernel \eqref{eq:dc_kernel_ct}, then the closed forms of  $\pmb{\Sigma}_{gy_0}(\tau)$  and $\pmb{\Sigma}_{y_0}$ exist as \eqref{eq:zoh_y_bl}. 
\end{Proposition}
In summary, we have shown that all three different combinations of the intersample and past behaviors have the closed forms of  $\pmb{\Sigma}_{gy_0}(\tau)$  and $\pmb{\Sigma}_{y_0}$.

\subsection{Closed Forms for Unknown Past Behavior}\label{se:cf_unknown}

In practice, the past behavior of the input signal is unknown, and then we need to make some assumptions on it and model the error in addition. 
As a result,  a transient model is considered in parallel with the system model, see e.g. \cite{Ljung04, PS12}, i.e., 
{\footnotesize\begin{align}\label{eq:extend_ir}
&y(t) = y_p(t)+ t_p(t)+ v(t),\ y_p(t) =  \int_0^\infty u_p(t-\tau)g(\tau)d\tau ,\\
&t_p(t) =   \int_0^\infty (u_p(t-\tau)-u(t-\tau))g(\tau)d\tau,\nonumber 
\end{align}}\normalsize
where $u_p(t)$ is the input with assumed past behavior,  $u(t)$ is the actual input, and  $t_p(t)$ is the transient model.

Here we employ the Bayesian framework of KRM for better exposition, corresponding to Lemma \ref{le:rels_time}. 
Let $t_p(t)$ and $g(\tau)$ are Gaussian processes and $v(t)$ is the white Gaussian noise.  This leads to a new MAP estimation as follows. 
\begin{Lemma}\label{le:ext_ir}
Consider \eqref{eq:extend_ir} with:
\begin{enumerate}
\item $g(\tau)$ is a zero-mean GP with covariance function $\kappa_g(\tau,\tau')$,
\item $t_g(t)$ is a zero-mean GP with covariance function $\kappa_t(t,t')$, independent with $g(\tau)$,
\item $v(t)$ is a white Gaussian noise with variance $\sigma^2$, independent with  $g(\tau)$ and $t_g(t)$.
\end{enumerate}
Given observation $\pmb y=[y(0),\cdots, y((N-1)T_s)]^T$, 
the MAP estimate of $g(\tau)$ is 
{\footnotesize\begin{align}
&\hat{g}(\tau)= \pmb  \Sigma_{gy_p}(\tau) (\pmb \Sigma_{y_p}+\kappa_{t}(\pmb{t},\pmb{t}^T)+ \sigma^2 I_N)^{-1}\pmb y, \label{eq:map_ext_ir} \\
&\left\{\begin{array}{lll}
\pmb \Sigma_{y_p}\!=\!
\int_{0}^\infty \int_{0}^\infty  u_p(\pmb t-\tau) u_p(\pmb t^T-\tau')\kappa_{g}(\tau,\tau')\,d\tau \,d \tau'  , \\
\pmb \Sigma_{gy_p}(\tau)=
 \int_0^\infty  u_p(\pmb t^T-\tau') \kappa_{g}(\tau,\tau')d\tau' .
 \end{array}\right.\nonumber
\end{align} }\normalsize
\end{Lemma}
\vspace{3mm}
It should be noted that $\kappa_t$ is a kernel in analogy with $\kappa_g$. The design of $\kappa_t$ can refer to \cite[Sec. 5.3]{LC16}, but it does not affect the existence of the closed form of the estimator \eqref{eq:map_ext_ir}.

Then we are interested in the closed forms of the new estimator \eqref{eq:map_ext_ir} (and also the new hyperparameter estimator), which depend on the closed form of $\pmb \Sigma_{y_p}$  and $\pmb \Sigma_{gy_p}(\tau)$. It should be noted that $\pmb \Sigma_{y_p}$  and $\pmb \Sigma_{gy_p}(\tau)$  take the same forms as \eqref{eq:map_y} with $u_p(t)$ having the known intersample and past behaviors. Consequently, all results can be found as special cases in Section \ref{se:cf_known}.

\section{Simulation}\label{se:sim}

We run numerical simulations to compare KRM with SRIVC and ML/PEM. 
Due to the limited space, we only consider the typical case in practice \cite{LJUNG09}: the intersample behavior is ZOH and the past behavior is unknown. 

\subsection{Data Generation}

The following setup is the same as tutorial 1 in CONTSID 7.4 toolbox \cite{GGMC21}. We consider the Rao–Garnier test system \cite{GY14}, whose transfer function is  
{\small\begin{align}\label{eq:rg_system}
G(s)= \frac{-6400s+ 1600}{s^4 + 6s^3 + 408^2 + 416s+1600}.
\end{align}}\normalsize
 The input is a pseudo-random binary signal (PRBS) generated by CONTSID function \texttt{prbs(10,7)} including  7167 samples in total, and the output is simulated by CONTSID function \texttt{simc()}, where the sampling interval is $T_s=0.01$, and the input is ZOH. Note that the past behavior is unknown. The additive noise is white Gaussian distributed with the signal-to-noise ratio (SNR) being 10 dB. Lastly, the 3001st to 4000th input and output are saved in one trail with $N=1000$, and 200 Monte Carlo (MC) trails are run with different inputs and noise realizations for one data bank. 

To enrich the benchmark, we also choose other values of $T_s$ and $N$. Specifically, we generate four data banks D1-D4: 
\begin{itemize}
\item[D1:] $T_s = 0.01$ and  $N$=1000, such that $N\cdot T_s=10$. 
\item[D2:] $T_s = 0.05$ and  $N$=200, such that $N\cdot T_s=10$. 
\item[D3:] $T_s = 0.1$ and  $N$=100, such that $N\cdot T_s=10$.
\item[D4:] $T_s = 0.1$ and  $N$=1000, such that $N\cdot T_s=100$. 
\end{itemize} 
It is shown in \cite{GMR03} that the normal sampling interval for system \eqref{eq:rg_system} is $T_s=0.05$, implying that $T_s = 0.01$ in D1 is over-sampled, and  $T_s = 0.1$ in D3/D4 is under-sampled.  
\subsection{Estimator Setup}
The tested estimators include 
\begin{itemize}
\item SRIVC, implemented by  \texttt{srivc()} in CONTSID Toolbox \cite{GGMC21}; 
\item PEM, implemented by \texttt{tfest()} in MATLAB System Identification Toolbox \cite{LJUNG09}; 
\item KRM, implemented by \eqref{eq:map_ext_ir} with $\pmb{\Sigma}_{y_p}$ and $\pmb{\Sigma}a_{gy_p}$ being \eqref{eq:zoh_zc_y_time} and kernel being the DC kernel \eqref{eq:dc_kernel_ct}.   
\end{itemize}
Specifically, SRIVC and ML/PEM need to select the orders of the numerator and denominator from $[1,9]$ and $[2, 10]$, respectively, with a constraint of the positive relative degree. For both SRIVC and ML/PEM, the model order selection is based on the Young information criterion (YIC) or Akaike information criterion (AIC), implemented by \texttt{selcstruc()} in CONTSID Toolbox with 'Algo' being 'srivc'. Besides, the true model orders are also used as the oracle (OR).
As for KRM, $\kappa_t$ is chosen to be $\alpha_t \kappa_g$ with $\alpha_t>0$ being an extra hyperparameter.  The hyperparameter estimation is implemented by the Empirical Bayes method with MATLAB function \texttt{MultiStart()} with  \texttt{fmincon()}, and the number of runs is 5 times the number of the hyperparameters. 

\subsection{Evaluation Criteria}
The evaluation criteria depend on the goal of the model, e.g. prediction and control design. For the goal of  prediction, the model is assessed by comparing the predicted output and the true noiseless output in the validation data: 
{\footnotesize \begin{align*}
 &\text{FIT}_y =100\bigg(1- \bigg[\frac{\sum\limits_{k=1}^{1000} |y_0(kTs)-\hat{y}(kT_s)|^2}{\sum\limits_{k=1}^{1000}|y_0(kT_s)-\bar{y}_0|^2}\bigg]^{\frac{1}{2}} \bigg), 
\end{align*}  }\normalsize
where $\bar{y}_0= \sum_{k=1}^{1000}y_0(kT_s)$,  and the true past input are provided for $\hat{y}$.
For the goal of  control design, the model is assessed by comparing  the estimated and true impulse response over-sampled at $\mathcal{X}_e=\{0.01/50:0.01/50: 10\}$:
{\footnotesize \begin{align*}
&\text{FIT}_g =100\bigg(1- \bigg[\frac{\sum\limits_{\tau\in \mathcal{X}_e} |g_0(\tau)-\hat{g}(\tau)|^2}{\sum\limits_{\tau\in \mathcal{X}_e}|g_0(\tau)-\bar{g}_0|^2} ]\bigg]^{\frac{1}{2}}\bigg),
\end{align*}}\normalsize
where  $\bar{g}_0= \sum_{\tau\in \mathcal{X}_e}g_0(\tau)$. 
The evaluation criteria are independent of the choices of $T_s$ and $N$.

\subsection{Results and Findings}

\begin{figure*}
\center
  \begin{minipage}{.48\textwidth}
\includegraphics[width=\linewidth]{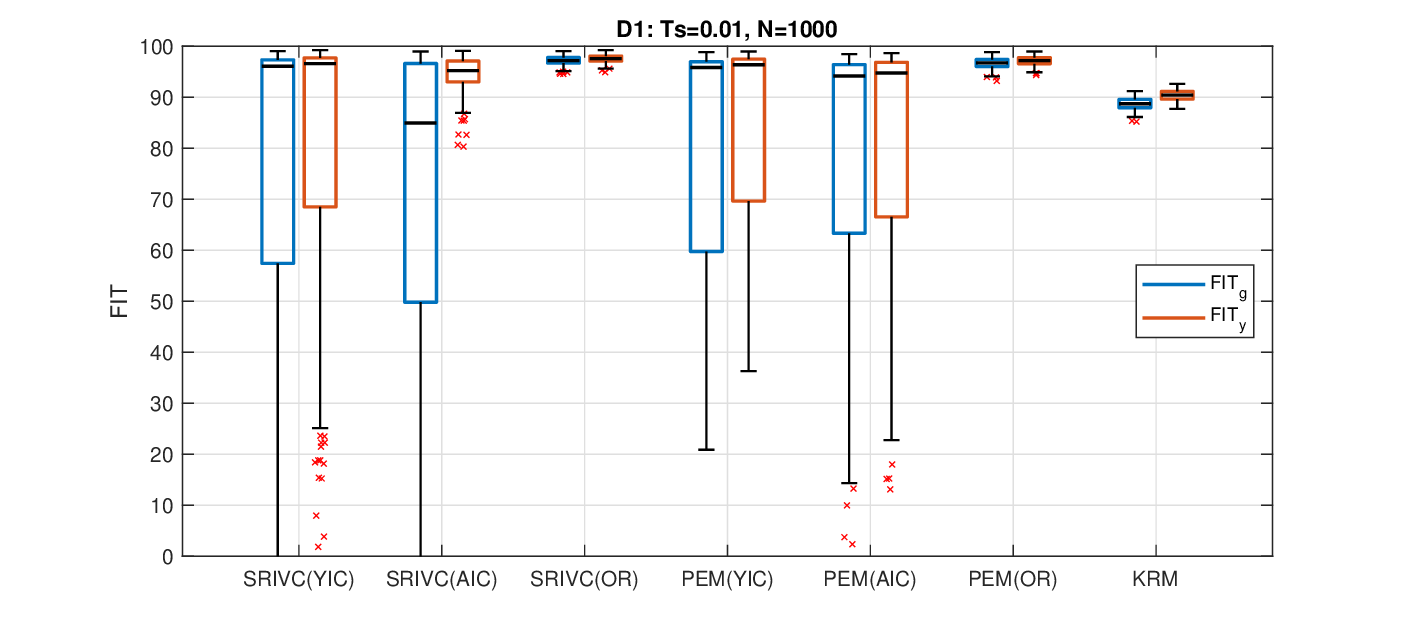}
\end{minipage}
  \begin{minipage}{.48\textwidth}
\includegraphics[width=\linewidth]{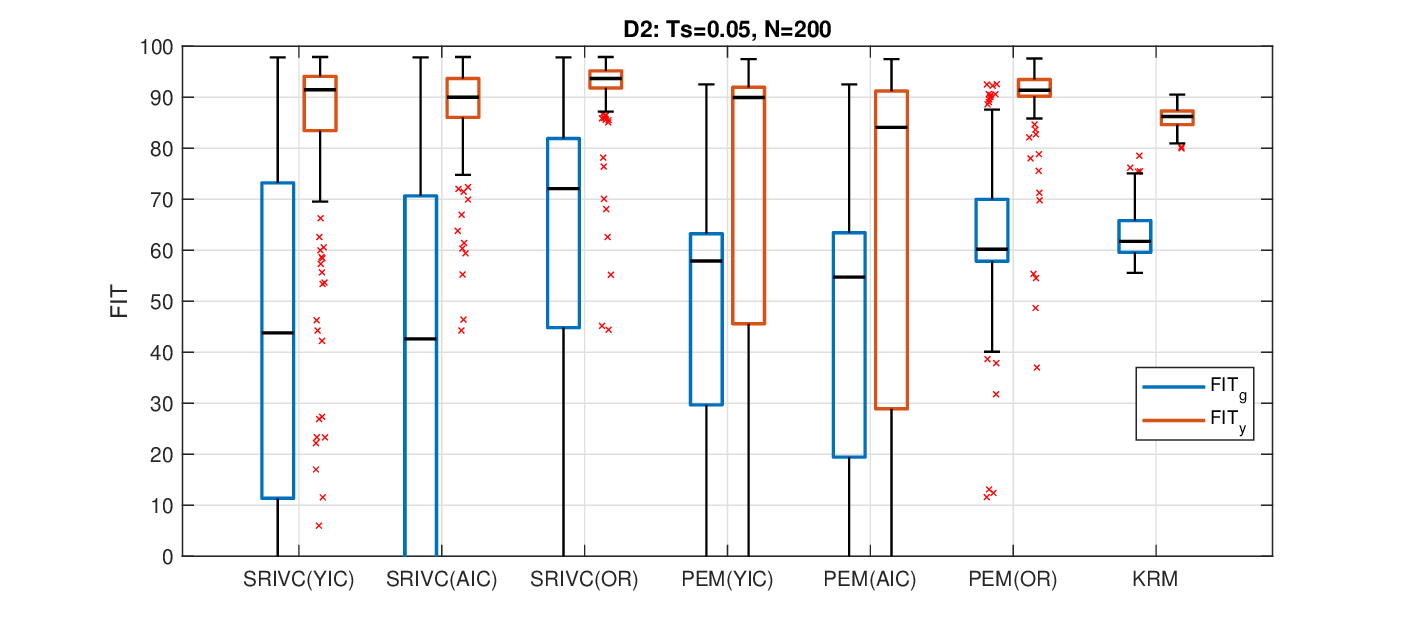}
\end{minipage}
  \begin{minipage}{.48\textwidth}
\includegraphics[width=\linewidth]{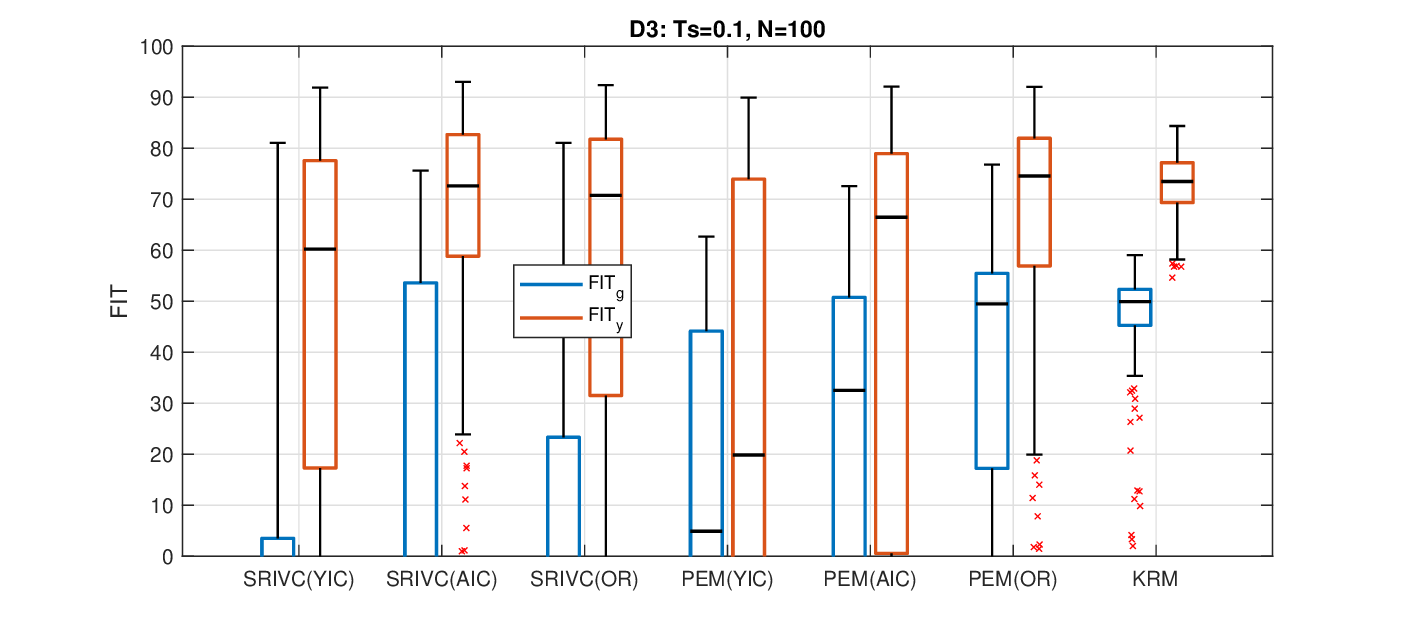}
\end{minipage}
  \begin{minipage}{.48\textwidth}
  \includegraphics[width=\linewidth]{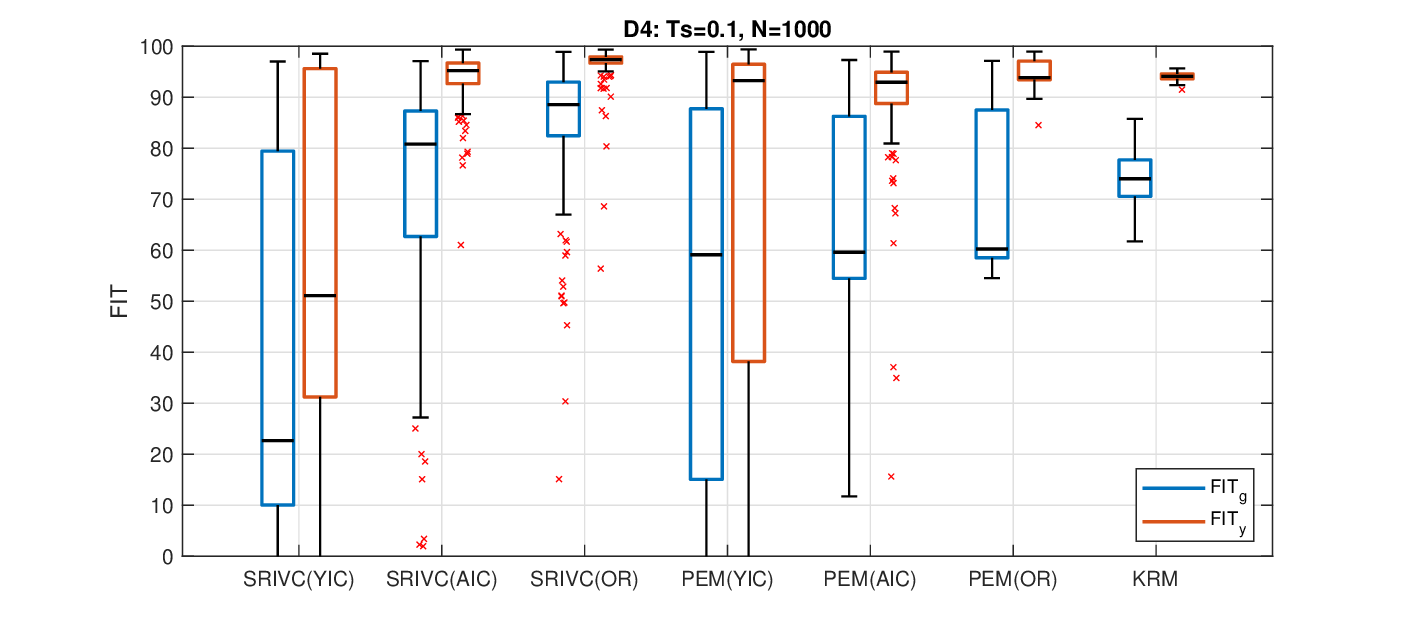}
\end{minipage}
\caption{The boxplots of FIT$_g$ (blue) and FIT$_y$ (red) for 200 MC trials. The mean and standard deviation refer to Table \ref{ta:fit}.  }  \label{fig:fit}
\end{figure*}

The boxplots of FIT$_g$ and FIT$_y$ are shown in Fig. \ref{fig:fit}, and the mean and standard deviation (std) of FIT$_g$ and FIT$_y$ are tabulated in Table \ref{ta:fit}.
The following tendency is consistent with our expectations:
\begin{itemize}
\item The larger $N$ improves both FIT$_g$ and FIT$_y$  in terms of the mean and std.  
\item The larger $T_s$  deteriorates FIT$_g$ in terms of the mean and std, while its influence on FIT$_y$ is relatively weak.
\end{itemize}
Moreover, we have the following observations on estimates: 
\begin{itemize}
\item for the estimator's robustness in terms of the std, KRM is more robust than SRIVC and ML/PEM, and they are equally robust in D1 (larger sample size $N$) with the true model order provided.
 \item for the estimator's accuracy in terms of the mean, KRM is more accurate than SRIVC and ML/PEM in D3 (smaller sample size $N$), and, if the true model order is unknown, then KRM is almost no worse than SRIVC and ML/PEM. 
\end{itemize} 
\begin{table}
    \caption{Mean and standard deviation (std) of FIT.}\label{ta:fit}
    \resizebox{\columnwidth}{!}{%
\begin{tabular}{cllll}
\toprule 
 \multicolumn{5}{c}{  $\text{Mean}_{\text{std}}$ of FIT$_g$}.\\
    \hline
Estimate & \quad D1& \quad  D2& \qquad D3& \quad  D4\\
\hline
KRIVC(YIC)&    $66.45_{54.97}$&   $30.10_{57.27}$& $-59.40_{82.11}$&$ 32.43_{49.77}$\\
KRIVC(AIC)&   $66.88_{42.04}$ &  $22.03_{67.27}$&$-47.05_{107.71}$& $69.19_{33.24}$ \\
KRIVC(OR)&   $\textbf{97.15}_{\textbf{0.88}}$& $ 53.68_{48.56}$& $-18.62_{60.76}$ &  $\textbf{84.07}_{17.21}$\\
PEM(YIC)& $69.29_{53.32}$& $ 34.07_{73.43}$&$-14.07_{68.51}$ & $43.63_{56.03}$\\
PEM(AIC)&    $58.40_{67.07}$ & $ 31.54_{64.83}$&\quad \ $5.88_{58.08}$& $61.06_{37.33}$\\
PEM(OR)&   $96.65_{ 1.01}$ & $ 59.72_{31.21}$&$\quad 25.07_{54.58}$&  $71.10_{17.14}$ \\
KRM&   $88.73_{1.16}$&   $\textbf{63.06}_{\textbf{4.62}}$&$\quad \textbf{45.34}_{\textbf{14.75}}$& $74.06_{\textbf{4.85}}$ \\
\hline
 \multicolumn{5}{c}{  $\text{Mean}_{\text{std}}$ of FIT$_y$}.\\
    \hline
Estimate & \quad D1& \quad  D2& \qquad D3& \quad  D4\\
\hline
SRIVC(YIC)& $69.47_{67.20}$&  $76.15_{38.45}$& \ \quad $38.33_{52.04}$  & $49.05_{51.34}$\\
SRIVC(AIC)& $93.67_{14.13}$& $85.30_{21.73}$&   \ \quad  $62.21_{38.52}$&$90.00_{27.69}$\\
SRIVC(OR)&  $\textbf{97.51}_{\textbf{0.79}}$&$ \textbf{92.04}_{7.05}$&  \ \quad $52.00_{61.79}$&$\textbf{96.57}_{4.05}$\\
PEM(YIC)&$70.17_{64.20}$& $ 49.00_{87.53}$&  $-122.25_{374.22}$ &$61.89_{57.06}$\\
PEM(AIC)&$56.49_{71.71}$&$ 40.33_{103.14}$& \,  $-12.58_{208.55}$ &$81.82_{37.85}$\\
PEM(OR)& $97.11_{0.87}$& $ 90.43_{7.10}$&   $\ \quad 56.37_{53.98}$& $94.93_{2.20}$\\
KRM&  $90.37_{1.04}$&   $86.03_{\textbf{1.96}}$&  $\ \quad \textbf{72.74}_{\textbf{6.06}}$&$94.10_{\textbf{0.69}}$\\
\bottomrule
  \end{tabular}}
\end{table}
\subsection{Conclusion}
This work proposes the kernel-based regularization method for continuous-time system identification from the sampled data.  The closed forms of the estimators are demonstrated when the input has certain combinations of the intersample and past behaviors. For practical use, the result can be extended to when the input has unknown past behavior.   According to the numerical simulations, the proposed method has demonstrated better robustness compared to the state-of-the-art methods, SRIVC and ML/PEM, and the proposed method also has better accuracy for the small sample size.   
\bibliographystyle{unsrt}
\bibliography{database}

\def\thesectiondis{\thesection.}                   % I.
\def\thesubsectiondis{\thesection.\arabic{subsection}.}          % B.
\def\thesubsubsectiondis{\thesubsection.\arabic{subsubsection}.}

\setcounter{subsection}{0}

\renewcommand{\thesection}{A}

\renewcommand{\theequation}{A.\arabic{equation}}
\setcounter{equation}{0}

\renewcommand{\thesubsection}{\thesection.\arabic{subsection}}

\renewcommand{\thesection}{A}
\renewcommand{\thesubsection}{A.\arabic{subsection}}

\section*{Appendix}

\subsection{Complementary Materials for Example \ref{ex:dc_kernel}}\label{ap:dc_example}
{\small
\begin{align*}
 \bullet\  &\kappa_{gg_d}^{\text{DC}} (\tau, s';\eta)=\left\{
     \begin{array}{llll} 
\text{EXP1},&  \text{if } s'-1<\tau/T_s\leq s'  \\
\text{EXP2},& \text{if }\tau/T_s> s',      \\
\text{EXP3},&  \text{if }\tau/T_s\leq  s'-1,  
   \end{array}\right.,\\
   \bullet\ & \kappa_{gg_p}^{\text{DC}}(\tau, \tau';\eta)=\frac{1}{1-e^{-(\alpha+ \beta) NT_s}} M_1\\
&  - \frac{1-e^{-(\alpha+\beta)NT_s C_1}}{1-e^{-(\alpha+ \beta) NT_s }}M_1+ \frac{1-e^{-(\alpha-\beta)NT_s C_1}}{1-e^{-(\alpha- \beta) NT_s}}M_2\\
&  M_1=\lambda e^{ -\alpha(\tau+ \tau')-\beta( \tau'-\tau) } ,\  M_2=\lambda e^{-\alpha(\tau+\tau')-\beta (\tau-\tau')}, \\
\bullet\ & \kappa_{gg_{dp}}^{\text{DC}}(\tau, s';\eta)\!=\!\left\{
     \begin{array}{llll} 
\text{EXP4},&  \text{if } s'-1<\text{mod}(\frac{\tau}{T_s},N)\leq s'  \\
\text{EXP5},& \text{otherwise},      \\ 
   \end{array}\right.,\\
 &\text{EXP1}=\frac{ \lambda}{\alpha^2 -\beta^2 }\bigg((\beta -\alpha ) e^{-\alpha(\tau+s'T_s)-\beta(s'T_s-\tau)}\nonumber\\
   &- 2 \beta  e^{-2\alpha \tau}+(\alpha +\beta ) e^{-\alpha (\tau+ (s'-1)T_s)- \beta(\tau-(s'-1)T_s)}\bigg),    \nonumber\\
   &\text{EXP2}=\frac{\lambda}{\alpha-\beta}\bigg(e^{-\alpha(\tau+(s'-1)T_s)}e^{-\beta|\tau-(s'-1)T_s|}\\
   &-e^{-\alpha(\tau+s'T_s)}e^{-\beta|\tau-s'T_s|}\bigg),\\
   &\text{EXP3}=\frac{\lambda}{\alpha+\beta}\bigg(e^{-\alpha(\tau+(s'-1)T_s)}e^{-\beta|\tau-(s'-1)T_s|}\\
   &-e^{-\alpha(\tau+s'T_s)}e^{-\beta|\tau-s'T_s|}\bigg),\\
   &\text{EXP4}= \lambda e^{-\alpha(\tau+ s'T_s)} \bigg(\frac{1}{1-e^{-(\alpha+\beta)NT_s }} S_5\\
   &- \frac{1-e^{-(\alpha+\beta)NT_s C_2}}{1-e^{-(\alpha+\beta) NT_s }}S_1 + \frac{1-e^{-(\alpha-\beta)NT_s C_2}}{1-e^{ -(\alpha-\beta) NT_s}}S_2\\
&- \frac{1-e^{-(\alpha+\beta)NT_s C_3}}{1-e^{-(\alpha+\beta) NT_s }}S_3  + \frac{1-e^{-(\alpha-\beta)NT_s C_3}}{1-e^{ -(\alpha-\beta) NT_s}}S_4 \bigg) \\
 &\text{EXP5}=  \lambda e^{-\alpha(\tau+ s'T_s)} \bigg(\frac{1}{1-e^{-(\alpha+\beta) N T_s}} S_5 \\
& - \frac{1-e^{-(\alpha+\beta)NT_s C_2}}{1-e^{-(\alpha +\beta) NT_s }}S_5+ \frac{1-e^{-(\alpha-\beta)NT_s C_3}}{1-e^{-(\alpha-\beta) NT_s}}S_6\bigg),
\end{align*}}\normalsize
{\small
\begin{align*}
&C_1= \text{ceil}[(\tau-\tau')/(NT_s)], C_2= \text{ceil}[(\tau/T_s-s')/N], \\
&C_3= \text{ceil}[(\tau/T_s-s'+1)/N], \\
&  S_1=\frac{e^{(\alpha+ \beta)T_s}- e^{-(\alpha+ \beta)(\text{mod}(\tau, NT_s)-s'T_s)}}{\alpha+ \beta} e^{-\beta( s'T_s-\tau) } ,\\
&  S_2= \frac{e^{(\alpha- \beta)T_s}- e^{-(\alpha- \beta)(\text{mod}(\tau, NT_s)-s'T_s)}}{\alpha- \beta}  e^{-\beta (\tau-s'T_s)}, \\
&  S_3=\frac{-1+ e^{-(\alpha+ \beta)(\text{mod}(\tau, NT_s)-s'T_s)}}{\alpha+ \beta}    e^{ -\beta( s'T_s-\tau) } ,\\
&  S_4=\frac{-1+ e^{-(\alpha- \beta)(\text{mod}(\tau, NT_s)-s'T_s)}}{\alpha- \beta}    e^{-\beta (\tau-s'T_s)}, \\
&  S_5\!=\!\frac{e^{(\alpha+ \beta)T_s}-1}{\alpha+ \beta} e^{-\beta( s'T_s-\tau) } , S_6\!=\!\frac{e^{(\alpha- \beta)T_s}-1}{\alpha- \beta} e^{-\beta (\tau-s'T_s)}.
\end{align*}\normalsize

\end{document}